\documentclass[journal=jacsat,manuscript=article]{achemso}
\setkeys{acs}{articletitle = true}

\def\bea{\begin{eqnarray}}
\def\eea{\end{eqnarray}}
\def\ben{\begin{equation}}
\def\een{\end{equation}}
\def\benu{\begin{enumerate}}
\def\enu{\end{enumerate}}

\def\sss{\scriptscriptstyle\rm}

\def\g{_\gamma}

\def\l{^\lambda}

\def\half{\frac{1}{2}}

\def\c{_{\sss C}}

\def\s{_{\sss S}}
\def\S{_{\sss S}}

\author{R. J. Magyar}  
\email{rjmagya@sandia.gov}
\affiliation{Center for Computing Research \& the Center for Integrated Nanotechnologies, Sandia National Laboratories, Albuquerque NM 87185}

\title{Time-dependent potential through an Ansatz for the Kohn-Sham orbitals}

\begin{document} 


\begin{abstract}
Given the time-evolution of an electron charge density, the local potential in Kohn-Sham time-dependent density functional theory (KS-TDDFT) can be modeled as a sum of instantaneous and dynamic
contributions by assuming a certain form of the time-dependent Kohn-Sham (TD-KS)  orbitals.  The instantaneous part is obtained numerically using methods from ground-state density functional theory (DFT) and the dynamic part is expressed in terms of a velocity potential that depends on the electron current density.  
The suggested form of the TD-KS orbitals satisfies several known constraints (orthonormality, N-representability, J-representability), and the domain of validity is shown to depend on the evolution of the instantaneous quantities.  Through this decomposition, we can relate time-dependent  and ground -state V-representability.     The resulting potentials are shown numerically to approximate the exact time-dependent Kohn-Sham potentials for a set of 3 non-singlet two-particle systems (a Kohn-mode, a Coulomb explosion, and a double quantum well) where the exact solutions and reference densities are  known or obtained through configuration interaction (CI) approaches.  
\end{abstract}




\section{Introduction}

Time-dependent density functional theory (TDDFT) within the Kohn-Sham representation \cite{RG84} enables the efficient computation of the dynamics of interacting electrons thereby  providing information for decision support in developing technologies such as photovoltaics, fuel cells, and qubits.  However, the accuracy of predictions is limited by the quality of approximations to the unknown time-dependent Kohn-Sham potential.    It is very rare that the exact potential is known but when available, comparisons between approximations and the exact can provide invaluable insight advancing the development of more reliable approximate potentials.    In this manuscript, we assume a form for the Kohn-Sham (KS) orbitals that is guaranteed to provide  1. a given time-dependent density from, for example, highly accurate quantum mechanics solutions, 2. the exact time-dependent current density, and 3. orbitals that are orthogonal to each other at all moments.   When the orbitals do have this form, the TD-KS potential can be decomposed into instantaneous and dynamic components that can be extracted independently using known techniques.   
The impact of these results is that a body of high precision quantum mechanical results can be made to provide exact TD-KS potentials, and the results can be compared to existing models and used to improve them.  This decomposition has implications for the structure of the theory by providing constraints on the resulting TD-KS potential.  

It is difficult to extract the TD-KS potential from high quality numeric results even approximately.   For singlet states of two electrons, the inversion can be done analytically.   Ullrich and others have examined the spin singlet state of 2 electrons in various external potentials noting the decomposition into adiabatically exact and non-adiabatic pieces \cite{U6,UT6}.  K{\"u}mmel and coworkers have studied exact models including 1D screened Coulomb systems to extract exact TDDFT results for spin-singlet, single KS orbital calculations \cite{LK5,WKL7,TGK8,TK9a,TK9b}.  It is also possible to invert exact solutions to study excited-states in the linear response formulation \cite{M9,TK14}.  We choose to explore the case when multiple Kohn-Sham orbitals of different characters are populated.  

The most straightforward approach to solve the inverse problem is to modify a trial potential self-consistently at each time step
so that an initial state propagates to a given density at the next time step; however, this approach is numerically demanding as the variations in the density are often quite small and smaller than significant variations in the KS orbitals \cite{RG12,HRC13}.  Numerically, this problem can be reformulated in the language of optimal control theory allowing inversion for a general number of particles but with a substantial computational overhead \cite{NRL13,JW16}.  These schemes share a high sensitivity to small variations of the exact density especially in the low density / large external potential regions.   We take another approach.  By restricting ourselves to a specific form of the orbitals, we find that we can describe the potential in 
terms of quantities that can be calculated more easily. 


Our work differs from earlier attempts to find the TD-KS potential in that we postulate an underlying relationship between time-dependent (TD) orbitals and instantaneous density through a Ansatz form of the orbitals.  This structure provides a direct route to the TD potential that is free of the two implementation problems suggested by Jensen \cite{JW16}.  These problems are the step by step integration and the division by infinitesimally small density contributions.   The former problem is mitigated by the inclusion of the velocity potential.     
The later problem is avoided  only in part since the instantaneously exact solutions can be solved in a way that avoids the inversion entirely; however, the problem does persist in the low density regime.  While it quite possible this Ansatz approach will not be valid in many cases of electron dynamics, the approach could provide potentials for analysis purposes as well as initial starting guesses for iterative processes.  

The introduction of hydrodynamic variables such as the local velocity field stems back to the early days of many-body theory and the relationship between effective potentials and the derivatives of the local velocity potential have also been suggested \cite{B52,DG82,B86}.  Here we show several important corollaries of this earlier work.  First, that the decomposition of the TD-KS potential breaks into instantaneous and dynamic pieces with the velocity field completely defining the later.  Second, that for many models problems this decomposition is a way to extract a TD-KS potential for more than two particles or for a two-particle non-singlet state.  

In the following, we restrict ourselves to the Schr{\"o}dinger equation for spin-less electrons in 1D for clarity and numerical convenience, and we use atomic units ($\hbar=m_e=e^2=1$), but the results can be generalized to the usual 3D case.

\section{Two Time-Dependent KS-like Systems}

We will describe the time-dependent electron density in two ways.  First, through the TD-KS formalism and then through what we define as \emph{the instantaneously exact KS formalism}.
The TD-KS equations for a set of orbitals are
\bea
\left(-\half\frac{d^2}{dx^2} +v\s(x,t)\right)\phi_i(x,t)=i\frac{\partial}{\partial t}\phi_i(x,t) , \;\;\; \sum_{i}^N \phi_i^\ast(x,t) \phi_i(x,t)=n(x,t). 
\label{tdkseq}
\eea
The KS orbitals, $\phi_i(x,t)$, are orthogonal to each other, and $n(x,t)$ is the time-dependent electron density.   $i$ runs from 1 to $N$, the number of electrons.  The TD-KS potential, $v\s(x,t)$, is assumed to exist and to be real.  In practice, the potential has contributions from the external potential, the Hartree term, and the exchange-correlation potential.  In this work, we intend to find an accurate representation for this potential as a whole.  While the Hartree term can be reliably constructed from an accurate density, the separation of the external potential and exchange-correlation potential from a reference density represents a further challenge.  If we generate our exact density from the solution of a many-body Schr{\"o}dinger equation, then we also know the external potential that is associated with the many-body density.  In this case, we can also find the exchange-correlation contribution to the potential. 

We define \emph{the instantaneously exact equation} as
\bea
\left(-\half\frac{d^2}{dx^2}  +v\s^{inst}(x,t)\right)\psi^{inst}_i(x,t)=\epsilon_i(t)\psi^{inst}(x,t), \;\;\; \sum_i^N \psi_i^{inst}(x,t)^2=n(x,t) 
\label{adkseq}
\eea
The instantaneous orbitals, $\psi^{inst}_i(x,t)$, are orthogonal to each other and real (in 1D) if they exist.  The $\epsilon_i(t)$ are the eigenvalues of the instantaneous Kohn-Sham Hamiltonian at time $t$.  The \emph{instantaneously exact  potential} is the potential that provides a stationary solution for the instantaneous density and can be obtained using methods similar to those suggested in Ref. \cite{LB94}.    Two important assumptions here are that the density is stationary v-representable and TD v-representable.   The designation \emph{stationary} is important here since \emph{the instantaneously exact potential} need not be the ground-state KS potential for the given density.  
\emph{It must however be an effective potential that produces a set of orbitals that reconstruct the given density, and the potential is continuously deformable from an earlier time's instantaneously exact potential}.  
It is possible that while tracking a time-dependent density, several potentials might provide a solution to the \emph{instantaneously exact equation} apparently violating a KS-like assumption.  However, the potential must be continuously deformable from the potential at the previous instance.  So if the system starts in a unique density functional ground-state, we will have a unique definition of the potential.    
We point out that it has been shown that a KS-like variational minimum can not always be found to be unique for excited state densities \cite{GB4,SHH6}.    It is unclear how this applies to the \emph{instantaneously exact} system as the assumption of being continuously deformable from an earlier time's potential may be sufficient to uniquely define this potential, and it is not required that the orbitals be occupied according to the lowest instantaneous eigenvalues as would be expected for the ground state problem.    In this paper we show some examples where this potential exists, but in general, the existence of the \emph{instantaneously exact potential} is a subject of needed inquiry. 
Considerable information about the system may also be extracted from the \emph{instantaneously exact} eigenvalues, $\epsilon_i(t)$.  If the state is stationary ground-state, these are the usual Kohn-Sham eigenvalues, but for a general density these provide additional information about the dynamics of the system.  In cases, where the eigenvalues cross in time, we might expect the dynamics to change character, where in a wave-function based theory, the single determinantal picture might be replaced by a multi-reference one.   

\emph{Key Ansatz}:  The TD-KS orbitals can be directly expressed in terms of the instantaneously exact orbitals and can be  written as
\bea
\phi_i(x,t)=   \exp\left(i S(x,t)- i \int_{t_0}^{t}d\tau\; \epsilon_i(\tau) \right) \psi^{inst}_i(x,t)
\label{guess}
\eea
where the modulus squared sum of these orbitals reproduces the density at a given snapshot in time and remains orthonormal,  
while the paramagnetic current, 
\bea
J(x,t)=-\frac{i}{2}\sum_{i=1}^{N} \left( \phi_i^\ast(x,t) \frac{d}{dx}  \phi_i(x,t)-  (\frac{d}{dx} \phi_i(x,t))^\ast \phi_i(x,t) \right),
\label{thecurrent}
\eea
satisfies the continuity equation.   Note that the inclusion of the position dependent phase in the wave-function can be compared to early work by Bohm \cite{B52} but the interpretation here is quite different.    The position-dependent phase can be related to the current, Eq. \ref{thecurrent}.  Conservation of electron charge requires that 
$\frac{\partial}{\partial t} n(x,t) + \frac{d}{dx} J(x,t) = 0$
where $n(x,t)$ is the time-dependent density and the local velocity field is $V(x,t)=J(x,t)/n(x,t)$.  
We will show the velocity potential, $S$, in our Ansatz Eq. \ref{guess} can be related to the current through $V(x,t)=\frac{d}{dx} S(x,t)= J(x,t)/n(x,t)$.  
Note that the time derivative of the velocity field $\dot{V} = 0 $ in the static limit.  For example if globally $\dot{V}$ vanishes, then we have either a stationary state at rest or a steady translational motion of a stationary distribution.  


To find the potential, we invert Eq. \ref{tdkseq},
\bea
v\s(x,t)=\half\frac{d^2}{dx^2} \phi_i(x,t) / \phi_i(x,t) +i\frac{\partial}{\partial t}\phi_i(x,t) / \phi_i(x,t),
\eea
and insert Eq. \ref{guess} for a TD-KS orbital, $\phi_i(x)$.   
The TD-KS potential is then
\bea
v\s(x,t)=
\half \frac{d^2}{dx^2}  \psi_i^{inst}(x,t) / \psi_i^{inst}(x,t)
-\half\left(\frac{d}{dx} S(x,t) \right) ^2
-\frac{\partial}{\partial t} S(x,t) + \epsilon_i(t)
\nonumber \\ 
+\frac{i}{2}\left(
\frac{d^2}{dx^2} S(x,t) 
+ 2 \frac{d}{dx} S(x,t) \frac{d}{dx} \psi_i^{inst}(x,t) / \psi_i^{inst}(x,t)
+2 \frac{\partial}{\partial t} \left(  \ln \psi^{inst}_i(x,t) \right) 
\right)  \label{vsham}
\eea
By inverting the \emph{instantaneously exact} KS-equation Eq. \ref{adkseq}, we have 
\bea
\frac{d^2}{dx^2} \psi_i^{inst}(x,t) / \psi_i^{inst}(x,t)  = 2 (v\s^{inst}(x,t) -\epsilon_i(t)). \label{adiainvert} 
\eea
For Eq. \ref{vsham} to describe a physically meaningful TD-KS potential, the imaginary part of the right-hand side must vanish.  Using the total derivative operator $\frac{D}{Dt}=\frac{\partial }{\partial t}+\frac{d}{dx} S \frac{d}{dx}$, we express the constraint as 
\bea
\frac{D}{Dt} \psi_i^{inst}(x,t) = -\half \frac{d^2}{dx^2} S(x,t) \psi_i^{inst}(x,t). \label{rhototd}
\eea
This condition can be tested at each time step since all the required quantities are available from the instantaneous calculation if the instantaneous solution exists.  Note that the result is consistent with the continuity equation for the charge density.  Using the total derivative notation, 
\bea
\frac{D}{Dt} n(x,t) = -n(x,t) \left(\frac{d}{dx} v(x,t) \right) = -n(x,t) \frac{d^2}{dx^2} S(x,t) . \label{continuity}
\eea
$n$ can be written $n(x,t)=\sum_i^N \psi_i^{inst}(x,t)^2$.   So we find a dynamic equation for the instantaneously exact orbitals,
\bea
\sum_i^N \psi_i^{inst}(x,t) \left(\frac{D}{Dt} + \half \frac{d^2}{dx^2} S(x,t) \right) \psi_i^{inst}(x,t) =0
\eea
Plug Eq. \ref{adiainvert} into Eq. \ref{vsham} and take the real part. 
The TD-KS potential is thus
\bea
v_S(x,t)=v_S^{inst}[n](x)|_{n=n(x,t)} + v_S^{dyn}(S,\dot{S})(x,t)
\label{vstdks}
\eea
The dynamic contribution to the potential is a \emph{function} of the velocity potential,
\bea
v\s^{dyn}(x,t)
=
-\frac{\partial}{\partial t} S(x,t)  - \frac{1}{2}\left(\frac{d}{dx} S(x,t) \right)^2.
\label{getvnad}
\eea
This result is general and does not require the solution to be a single electron or a 2-electron spin-singlet state.  We will demonstrate that for three non-trivial examples, this Ansatz holds within our numerical fidelity.

\section{Exact Results}
 
The \emph{exact} time-dependent densities are obtained by theory or by diagonalization of the few-body Hamiltonian  within a local orbital \emph{many-particle} basis.  
The initial state can be obtained to arbitrary accuracy by augmenting the basis set until convergence.   Our single particle basis set is the the set of solutions to the 1D non-interacting harmonic oscillator, and the 2-body basis is constructed as the set of antisymmetric products of one-body basis functions.  In the CI solutions, the entire product basis set up to a given upper limit in energy in the single particle basis is considered.  The dimension of the matrix being diagonalized is $\half(M^2-M)$ where M is the number of basis functions.
The bottleneck for the CI calculation are the calculation and storage of the 4-point integrals which scales as $M^4$.  The use of symmetries and fast identification of vanishing terms can reduce this cost substantially; nevertheless, the entire calculation is often limited to single particle bases of about 100 terms.   All matrix elements are evaluated analytically. 
The 4-point integrals can be reduces to products of special functions using elementary rules and the identity $\int_{-\infty}^\infty du\; \exp(-1/2 u^2) / \sqrt{\epsilon^2+u^2} =\exp(-\epsilon^2/4)K_0(\epsilon^2/4)$.
Time-dependent integration is performed using Crank-Nicolson integration in the two-particle basis.  The time steps are on the order of 0.01 in atomic units chosen small enough to maintain stable stationary state solutions of the many-body problem.  No predictor-corrector step is needed as the many-body Hamiltonian is known explicitly at all times.

In general, computing of the velocity potential on a discrete real space grid can be performed in a variety of  techniques for partial differential equations.    However, in 1D, we solve the continuity equation explicitly
\bea
S(x,t)=\int_{-\infty}^x dx'\; \frac{1}{n(x',t)} \left[ \int_{-\infty}^{x'} dx''\;  \frac{\partial}{\partial t} n(x'',t)\right] 
\label{sin1d}
\eea
assuming the charge density and current vanish sufficiently fast with $x$ that the spatial derivative of $S$ vanishes far from the region of interest.   Solving Eq. \ref{sin1d} is not ideal since one over the density is expected to diverge in regions far removed from significant density.  However, when analytic results are known, or when asymptotic limits for $S$ are available, this is an efficient way to obtain $S$.  

We investigate 3 systems.  
First, we consider the collective motion of a density distribution in a harmonic well.  This system yields explicit analytic results  and has been studied 
in the context of non-adiabatic functionals \cite{D94,V95}.  Since exact results are known here, the model offers a trial case for the methods proposed.  
The second system models a \emph{Coulomb} explosion.  Two initially harmonically confined electrons are released when the confinement potential is suddenly removed.  The system admits an exact solution in the non-interacting limit and is important for radiation damage in solids. 
The third system is an $x^4$ well with two local minima chosen to locally mimic the harmonic well of system 1.  The ground state localizes charge on discrete sites, but a linear perturbation can transfer charge from one site to the other.  Thus, this model allows us to examine an exact solution in a case when the derivative discontinuity should matter.  Additionally, we would 
transition from a system that goes from double reference to single reference directly probing some of the most challenging aspects of TDDFT. 



\begin{figure}[t]
\centering
\includegraphics[width=6in]{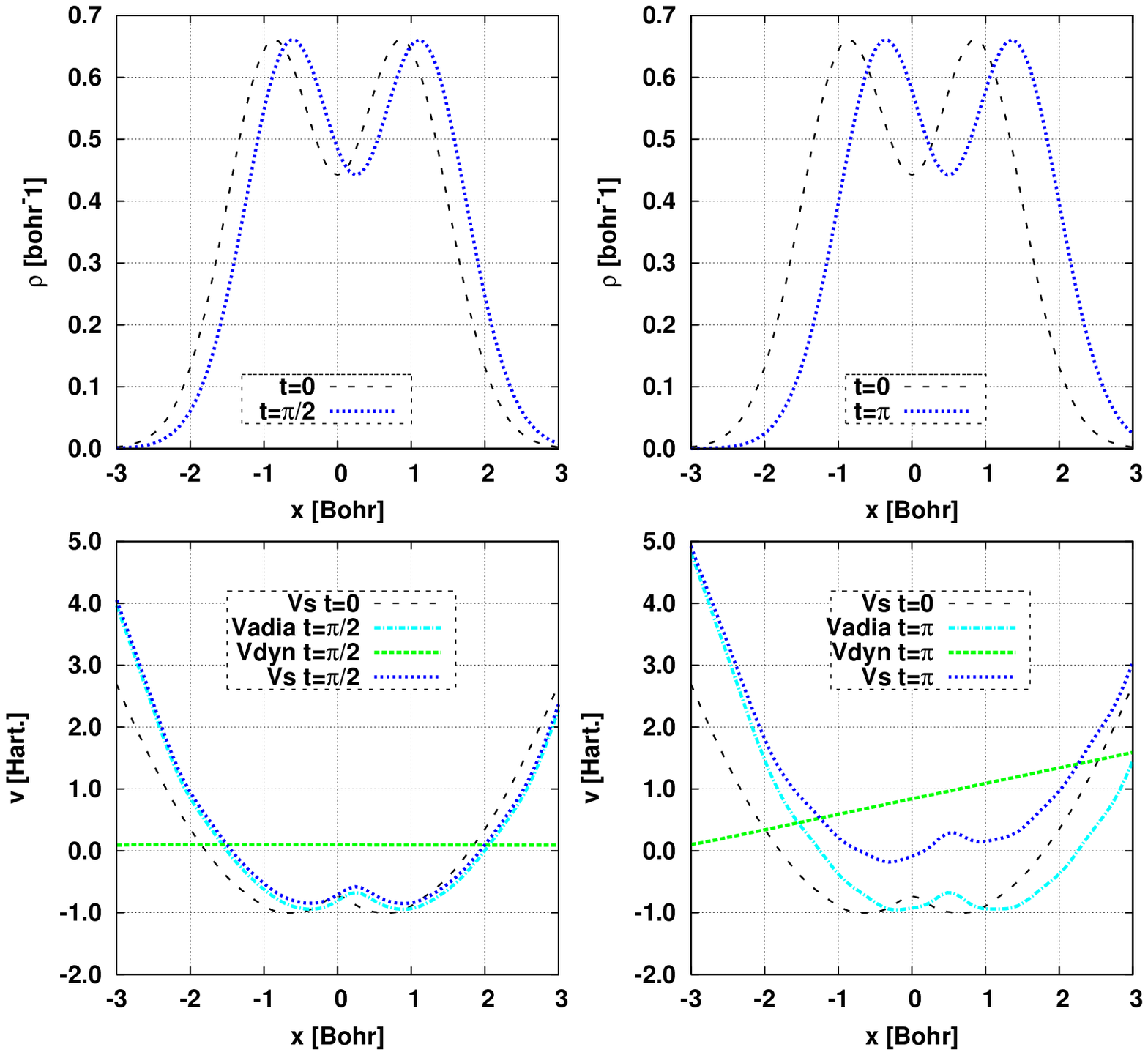}
\caption{Harmonic Potential Motion:
Time-evolution in a harmonic ($\omega=1$) potential after the the instantaneous application of an electric field (linear potential $\kappa=1/2$).  The density profile changes position according to simple harmonic motion but does not change shape according the the harmonic potential theorem.    Top left: density at initial time (long-dashed) and at maximum velocity (short-dashed) ($t=\frac{\pi}{2}$).  Top right: density at initial time and at the other turning point ($t=\pi$).  Bottom left: the time-dependent KS potential  ($v\s$), the instantaneous potential ($v_{inst}$), the dynamic potential ($v_{dyn}$) at initial and maximum velocity times.  Bottom right: same but for the other turning point.}
\label{f:x2} 
\end{figure} 

\begin{figure}[t]
\centering
\includegraphics[width=6in]{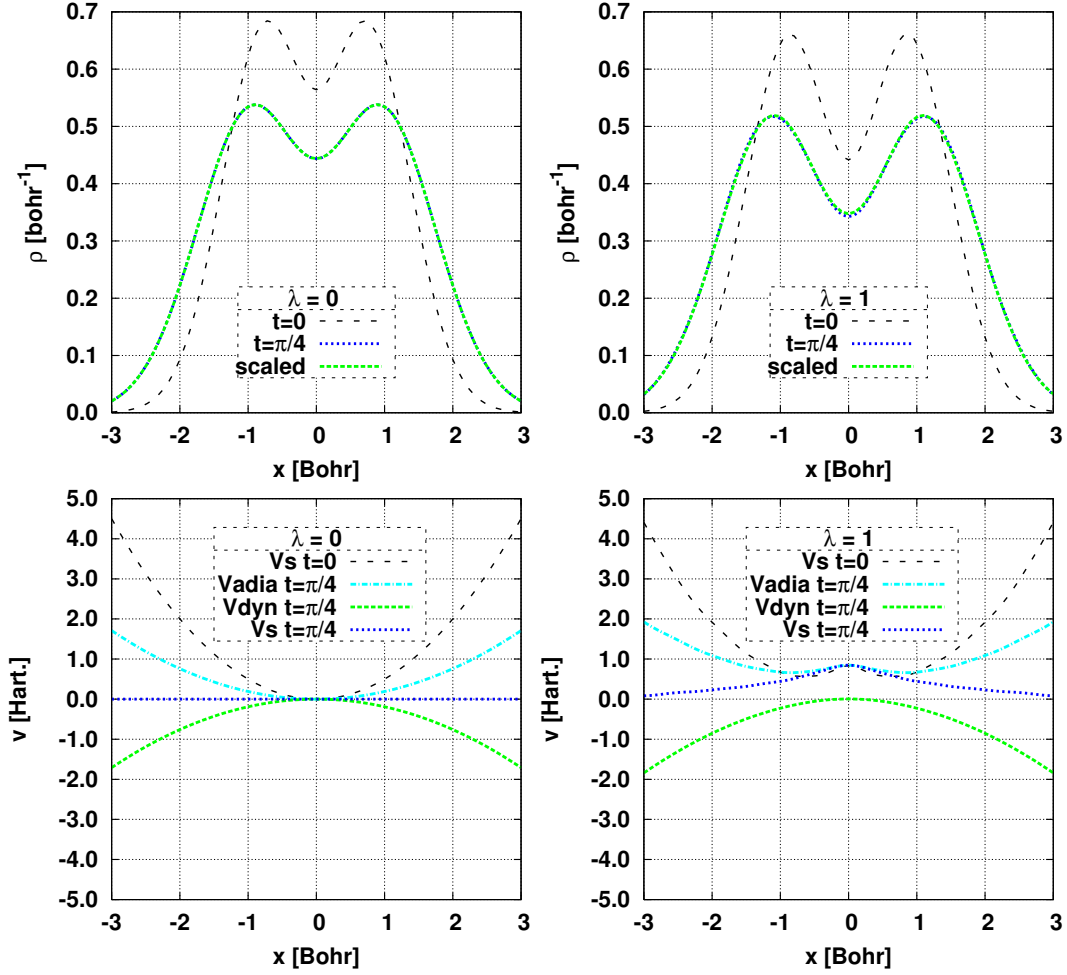}
\caption{The \emph{Coulomb} Explosion: Time-evolution after the instantaneous annihilation of a harmonic ($\omega=1$)  confinement potential.  The plots on the left are for $\lambda=0$ the non-interacting and analytically solvable system, and on the right, the interacting-system $\lambda=1$.  The densities are shown on the top.  The dashed line is the initial density, the short dashed blue line is the exact result, and the medium green-dashed line is the result of our TDDFT inversion.  
On the bottom are shown the corresponding potentials  $v\s$,$v_{inst}$,$v_{dyn}$ (dotted-blue, turquoise-dashed, short-green-dashed), the long black dashed in the original $v\s$.}
\label{f:ce} 
\end{figure} 



\begin{figure}[t]
\centering
\includegraphics[width=6in]{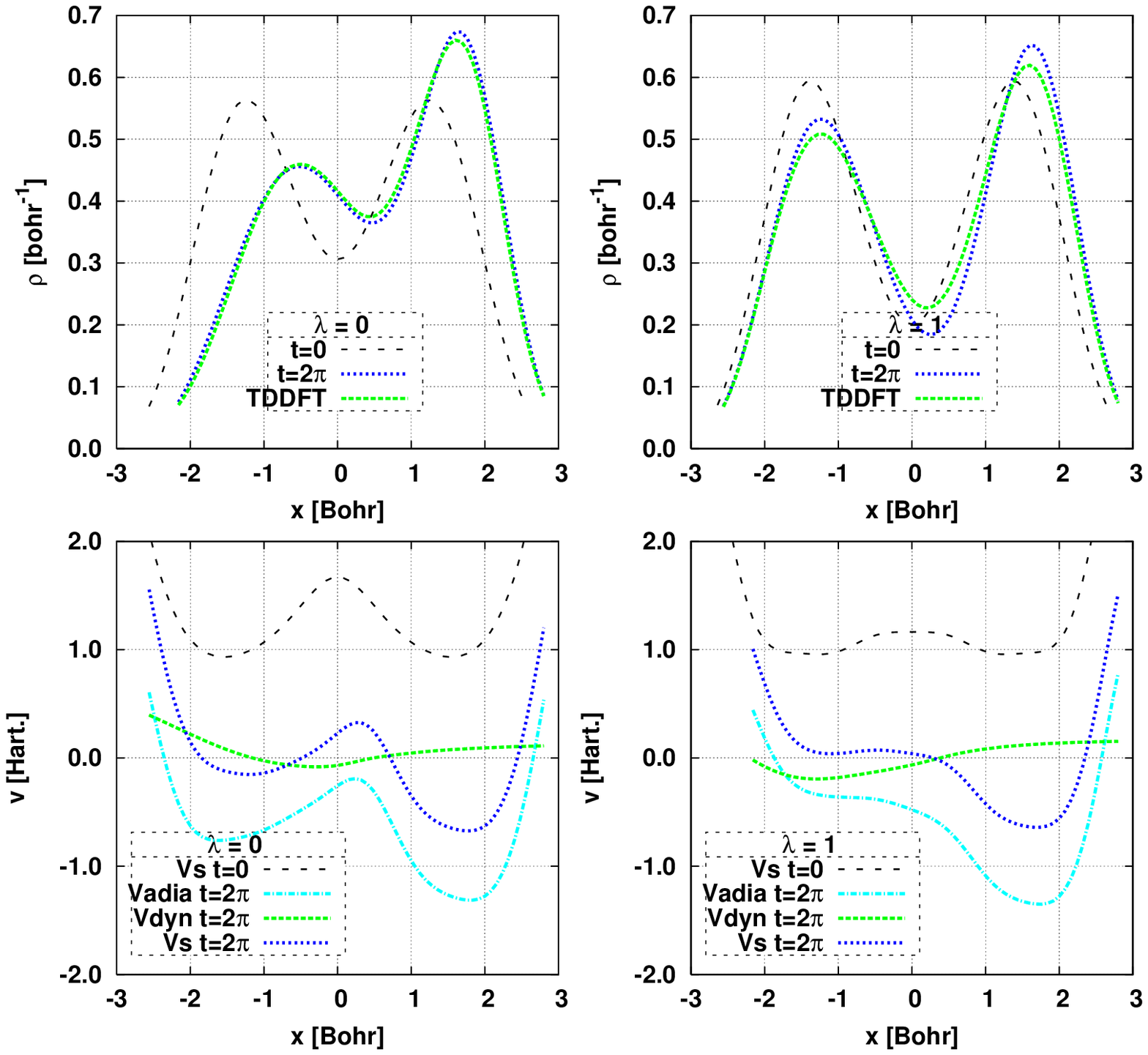}
\caption{Charge transfer excitation: Time-evolution of two electrons in a 1D double-well potential ($x_0=\sqrt{2}$ \& $\omega=1$ ) after being perturbed by an electric field.  The left side shows the density and potential for non-interacting electrons ($\lambda=0$).  The right side shows the density and potential for interacting ones ($\lambda=1$).  The densities and potentials are shown for the initial conditions (long-dashed). For the time-evolved densities, the top plots present the exact time-evolved results (short-blue dashed), and the results of the TD inversion scheme noted in the text (medium length, green dashed).  In the lower plots, we have the instantaneous contribution (alternating length, blue dashes), the dynamic contribution (medium length, green dashes), and the total potential (short, dark blue dashes).}
\label{f:ct} 
\end{figure} 





We chose to model the Coulomb interaction in 1D using the soft form $v(x)=\lambda/\sqrt{x^2+\epsilon^2}$ with $\lambda=1$ and $\epsilon^2=0.01$.  These parameters differ from those chosen in other work \cite{LK5}.  The advantages of this interaction potential choice are the numerical efficiency by which integrals can  be solved and the long ranged nature of the interaction; however, the potential violates expected scaling laws for a Coulomb interaction.  The small $\epsilon$ is deliberately chosen to enhance the short range effects of the interaction that are likely more accurately described using our local basis set.

\section{Example 1: Hooke's Atom}

The one-dimensional analog of Hooke's atom has two electrons occupying a harmonic well-potential and interacting through the screened Coulomb interaction.  The related Hooke's atom model in 3D has been used extensively to study exchange-correlation in TDDFT and the model is a natural starting place for study since  it admits exact solutions.  Dobson used the Harmonic potential theorem to study the adiabatic local density approximation \cite{D94} providing results exploited heavily here to validate our calculations and to illustrate the generality of the theorem.  A linear term simulating a constant electric field is applied instantaneously at $t=0$.
\bea
v_{ext.}(x,t)
 = \left\{
\begin{array}{lr} 
\half\omega^2 x^2,&     t<0 \\
\half\omega^2 x^2+\kappa x ,&   t \geq 0 \\
\end{array}\right.
\eea
The central peak of the density is at $X(t)=\frac{\kappa}{\omega^2}(1-\cos(\omega t))$.  The exact density of a many-fermion system time evolves according to $n(x,t)=n_0(x-X(t))$ where $n_0$ is the initial density profile either interacting or not.   The velocity potential is $S(x,t)=\frac{\kappa}{\omega} x\sin(\omega t)
$.  Note that since $\frac{d^2}{dx^2} S(x,t)=0$ this is incompressible flow of the electron liquid.  
The instantaneously exact KS potential,
\bea
v_{ks}^{inst}(x,t)=\frac{\omega^2}{2} (x-X(t))^2+v_{HXC}^0(x-X(t)) \\
= 
\frac{\omega^2}{2} x^2
-\omega^2  x X(t)
+v_{HXC}^0(x-X(t))
+\frac{\omega^2}{2} X(t)^2,
\eea
is just the translated potential since in a Kohn-model oscillation only the center of charge changes position but the distribution retains the same shape.  Note that the final term is just a time dependent constant shift of the potential and does not affect the dynamics of the density.
The exact TD-KS potential is \cite{D94}
\bea
v_{ks}(x,t)=\frac{\omega^2}{2} x^2+v_{HXC}^0(x-X(t))-\kappa x
\eea
Plugging $S$ into Eq. \ref{getvnad},
the dynamic potential is
\bea
v_{ks}^{dyn}(x,t)=\omega^2 x X(t)-\kappa x
\eea
which can be added to the instantaneous potential to provide the exact KS potential.  
Figure \ref{f:x2} shows the time-evolution in a harmonic ($\omega=1$) potential after the the instantaneous application of an electric field (linear potential $\kappa=1/2$).  The density profile changes position according to simple harmonic motion but does not change shape according to the harmonic potential theorem.    The plot is broken into quadrants with the densities at time zero (long-dashed)  and at the maximum velocity (top left, short-dashed) ($t=\frac{\pi}{2}$) and opposite turning points  (top right, short-dashed)($t=\pi$).   The bottom quadrants present elements of the effective time-dependent potential.  The time-dependent KS potential  ($v\s$), the instantaneous potential ($v_{inst}$), the dynamic potential ($v_{dyn}$) at initial and maximum velocity times and for the other turning point.   In this case, the instantaneous potential follows along with the density effectively maintaining the same shape.  The dynamic potential acts as a linear driving potential enforcing simple Harmonic motion.  At the maximum velocity point, there is interestingly no dynamic potential in this case; instead, the density's inertia drives the continued motion through the centroid.   These images are consistent with Dobson's earlier work, but demonstrate that this decomposition as implemented provides numerically expected results in a limit where they are available.


\section{Example 2: A Dispersing Wave-packet}

The second case to consider is the released and dispersing wave-packet. In the interacting case, this can be thought of as a Coulomb explosion,
\bea
v_{ext.}(x,t)
 = 
\left\{
\begin{array}{lr} 
\frac{\omega^2}{2} x^2,&     t<0 \\
0,&   t>0 \\
\end{array}\right.
\eea
The non-interacting density can be shown to time evolve according to  
\bea
n(x,t)=1/\sqrt{1+\omega^2 t^2} \; n_0(x/\sqrt{1+\omega^2t^2})
\eea
and $S(x,t)=\frac{\omega^2 x^2 t} {2(1+\omega^2 t^2)}$ \cite{M97}.  The time-evolved density is identical to the ground-state density with a different constant $\Omega(t)=\omega/(1+\omega^2 t^2)$.  Our numerical studies suggest that the interacting density obeys a similar scaling to high accuracy.  The figure \ref{f:ce} shows the time-evolution after the instantaneous annihilation of a harmonic ($\omega=1$)  confinement potential.  The plots on the left are for $\lambda=0$ the non-interacting and analytically solvable system, and on the right, the interacting-system $\lambda=1$.  The densities are shown on the top.  The dashed line is the initial density, the short-dashed blue line is the exact result, and the medium length, green-dashed line is the result of our TDDFT inversion.  
On the bottom are shown the corresponding potentials  $v\s$, $v_{inst}$, $v_{dyn}$ (dotted-blue, turquoise-dashed, short-green-dashed), the long black dashed in the original $v\s$.  As in the previous case, we find for the non-interacting case that the dynamic potential exactly cancels the instantaneously exact one correctly  predicting the total potential to vanish.  This is also calculated numerically for the \emph{interacting} case.    For the non-interacting case,
$v\s^{inst}=\frac{\omega^2}{2}\frac{x^2}{1+\omega^2 t^2}$
and
$v\s^{dyn}=-\frac{\omega^2}{2}\frac{x^2}{1+\omega^2 t^2}$.  
A Harmonic density expands, to a good approximation, according to a scaling formula even when interactions are present.  If this decomposition holds more generally for non-harmonic external potentials and in 3D, a new universal constraint on the TD-KS wave-functions must hold, the spatially-varying phase on the KS orbitals must be the same for all occupied orbitals. 
Plot \ref{f:ce} represents a \emph{Coulomb} explosion, in this case, because we use a screened-Coulomb interaction, this is more properly labeled an interaction explosion.  

\section{Example 3: The Double Well}

To illustrate utility, we apply this to 1D problem that does not admit explicit solutions even in the non-interacting case.  The problem is the 1D double well with an linear potential designed to shuffle one electron from the left well to the right well sharing the right site.  This model represents a charge-transfer event,
\bea
v_{ext.}(x,t)
 = 
 \left\{
\begin{array}{lr} 
\frac{\omega^2}{8 x_0^2} x^4 - \frac{\omega^2}{4} x^2            ,&     t<0 \\
\frac{\omega^2}{8 x_0^2} x^4 - \frac{\omega^2}{4} x^2  +\kappa x ,&   t>0 \\
\end{array}
\right.
\eea
with $\pm x_0$ being the position of the 2 wells and $\omega$ being the approximate angular frequency of the lowest state in each well.  Initially, each electron sits predominantly  in a separate quantum well.  The perturbing potential is designed to transform the initial double well into a single doubly occupied well.  We have simulated the dynamics of 2 cases. The first is a set of non-interacting electrons.  The second is a pair of interacting Fermi electrons.  We present results at an instant of time when a significant fraction of the charge has transferred to the combined-well.  

Figure \ref{f:ct} shows a charge transfer excitation: Time-evolution of two electrons in a 1D double-well potential ($x_0=\sqrt{2}$ \& $\omega=1$ ) after being perturbed by an electric field.  The left side shows the density and potential for non-interacting electrons ($\lambda=0$).  The right side shows the density and potential for interacting ones ($\lambda=1$).  The densities and potentials are shown for the initial conditions (long-dashed). For the time-evolved densities, the top plots present the exact time-evolved results (short-blue dashed), and the results of the TD inversion scheme noted in the text (medium green dashed).  In the lower plots, we have the instantaneous contribution (alternating blue dashes), the dynamic contribution (medium green dashes), and the total potential (short dark blue dashes).

This process moves fastest for the non-interacting case; for in the interacting case, Coulomb interaction discourages the free flow of electrons to the occupied site.  In both cases, the proposed decomposition provides a TD-KS potentials that accurately describes the dynamics of the electrons.   It is important to note the step in the \emph{dynamic potential}.   This is shown in Fig. \ref{f:ct} bottom plots.  The green-medium dashed line exhibits a dynamically changing step around $x=0$ ensuring that 2 wells have a correct chemical potential difference.    
The results manifests in the dynamic contribution to the TD-KS potential.  We show both the exact time-evolved densities and the results of our Ansatz for the TD-KS orbitals.  While the suggested approach proves accurate results for this highly non-trivial example, the density does drift from the exact one.  This is in part due to the difficulty in obtaining the instantaneously exact  potential in the asymptotic regimes through Eq. \ref{adkseq}.  The same can be said for the numeric calculation of the velocity potential.  Ultimately, the numeric challenges limit the general applicability.  However, the underlying structure of the TD-KS orbitals when true provides an strong insight into the requirements that need to be build into TD-KS potentials. 

\section{Conclusions}

In this paper for three 1D systems, the exact TD Kohn-Sham potential can be decomposed into instantaneous and dynamic contributions with variational theory providing the former and fluid-dynamic-like equations providing the latter.   This decomposition, when true, provides a route to new functionals by combining state-of-the-art ground-state functionals with approximations from fluid dynamics.

Additionally, the TD-KS orbitals can be related to the \emph{instantaneously exact} orbitals at each time step.  This reduces the problem of TD V-representability to a form of ground-state V-representability.  There is no paradox with initial-state dependence as the \emph{instantaneously exact potential} is assumed to be continuously deformable in time from the initial state, but the history dependence is contained somewhat paradoxically in the ground-state-like term.  It is expected that the consequences of this connection are that the \emph{instantaneous} orbitals are not necessarily filled in the order of increasing instantaneous eigenvalues.  

One of the potential drawbacks for using this Ansatz and decomposition to determine time-dependent potentials is that the instantaneous inversion requires the repeated solution of an eigenvalue problem scaling poorly with system system.  This problem is somewhat mitigated by the assumption that the instantaneous potential various continuously in time thereby providing a convenient starting place for the instantaneous solutions.   
However, the ability to calculate the instantaneous solutions will eventually limit the applicability of this method to systems below some critical size.   A related objection is that the solution of Eq. \ref{sin1d} involves a double integral and inverses of exponentially small densities.  This is not a sever  problem, and Eq. \ref{sin1d} can be solved using numerical methods for partial differential equations.  In greater than 1D, the use of Eq. \ref{sin1d} to find the velocity potential is problematic.  Instead direct solution of Eq. \ref{sin1d} is required to provide the velocity field.    In this case the dynamic potential will capture the singularities.  However, instantaneously exact V-representability will probably not hold for arbitrary time-evolving electron densities.  

In this paper, we have shown that an explicit representation of the time-dependent Kohn-Sham orbitals in terms of a position-dependent phase term and \emph{instantaneously exact orbitals} can provide an alternate scheme 
to construct the TDDFT potential from a target time-dependent density.  This approach avoids many of the instabilities that plague more direct inversion schemes.  However, there exists some limitation to this approach as the existence of the instantaneously exact potential is not always guaranteed, and even then, the instantaneous solution might still be numerically challenging.  Additionally, the scheme does not distinguish between external and exchange-correlation contributions to the potential.  If the driving potential for the exact density is known then the exchange-correlation potential may be extracted independently.  We plan to generalize this procedure to 3D and to explore more realistic Coulomb interacting potentials.

We would like to thank A.D. Baczewski, N. Maitra, and D. Jensen for insightful discussions.   This work was funded through the Sandia National Laboratories LDRD office under project 151362 in the Science of Extreme Environments research area.  Sandia National Laboratories is a multi-program laboratory managed and operated by Sandia Corporation, a wholly owned subsidiary of Lockheed Martin Corporation, for the U.S. Department of Energy'€™s National Nuclear Security Administration under contract DE-AC04-94AL85000.




 \end{document}